\documentstyle[buckow]{article}
\begin {document}
\topmargin -25mm
\begin{flushright}
\begin{large}
IHP-2000/07\\
ULB-TH/00-32\\
hep-th/0012203\\
\end{large}
\end{flushright}
\def\titleline{
Non--semisimple gaugings 
of $D=5~~{\cal N}=8$ Supergravity
}
\def\authors{
Laura Andrianopoli\1ad,  Francesco Cordaro\2ad, 
Pietro Fr\'e\3ad, Leonardo Gualtieri{\normalsize$^{4,5}$} 
}
\def\addresses{
\1ad Dipartimento di Fisica, Politecnico di Torino\\ 
corso Duca degli Abruzzi 24, I-10129 Torino, Italy\\
\2ad Instituut voor Theoretische Fysica - Katholieke Universiteit Leuven\\
Celestijnenlaan 200D B--3001 Leuven, Belgium\\
\3ad Dipartimento di Fisica Teorica, Universit\'a di Torino\\
via P. Giuria 1, I-10125 Torino\\
Istituto Nazionale di Fisica Nucleare (INFN) - Sezione di Torino,
Italy \\
\4ad Physique Th\'eorique et Math\'ematique, Universit\'e Libre de
Bruxelles\\ 
C.P. 231, B--1050, Bruxelles, Belgium \\
\5ad Centre Emile Borel, Institut Henri Poincar\'e\\ 
11 rue Pierre et Marie Curie, F- 75231 Paris Cedex 05, France
}
\def\abstracttext{
The recent construction of the non--semisimple gaugings of maximal 
$D=5$ supergravity is reported here. This construction is worked out 
in the so--called rheonomic approach, based on Free Differential Algebras 
and the solution of their Bianchi identities.
In this approach the dualization mechanism between one--forms and 
two--forms is more transparent. The lagrangian is unnecessary since the 
field equations follow from closure of the supersymmetry algebra.
These theories contain $12-r$ self--dual two--forms and $15+r$ gauge 
vectors, $r$ of which are abelian and neutral.
Such theories, whose existence has been proved and their supersymmetry 
algebra constructed, have potentially interesting properties in relation 
with domain wall solutions and the trapping of gravity.
}
\large
\makefront
\section{Introduction}
Gauged supergravity with a maximal compact group, 
${\cal G}= SO(6)$ in $D=5$
\cite{gunwar,PPV}, ${\cal G}=SO(8)$ in
$D=4$ \cite{dewitnicol} and ${\cal G}=USp(4)$ in $D=7$ 
\cite{Pernici:1984xx} has attracted much renewed attention in the last two
years because of the $AdS_{p+2}/CFT_{p+1}$ correspondence 
(for a general review see \cite{Aharony:2000ti} and references therein). 
Indeed the maximally supersymmetric vacuum of these gauged supergravities 
is the $AdS_D$ space and the compact gauge group ${\cal G}_{gauge}$ is 
the $\cal{R}$-symmetry of the corresponding maximally extended 
supersymmetry algebra.
However the compact gaugings are not the only ones for extended
supergravities. There exist also versions of these theories where the gauge
group  ${\cal G}_{gauge}$ is non--compact. Unitarity is preserved 
because in all possible extrema of the corresponding  scalar potential the
non--compact gauge symmetry is broken to some residual compact subgroup.
Furthermore, there are models in which the gauge group is non--semisimple.
\par
The non--semisimple gauged supergravities are  relevant for a close 
relative of the
$AdS/CFT$ correspondence, namely the
{\it Domain Wall/QFT correspondence} \cite{Boonstra:1999mp}
between gauged supergravities and quantum field theories realized on 
certain domain wall solutions of either string theory or M--theory. 
Recently, it has been shown \cite{Cvetic:2000pn} that these 
domain wall solutions are related with localization of gravity. 
So, non--semisimple gauged supergravities could be good candidates 
to accomodate the Randall Sundrum scenario 
\cite{Randalsud} into supergravity theories.
\par
Here we report the recent construction \cite{cso} of the non--semisimple 
gaugings for $D=5$ maximal supergravity. 
In the case of $D=4$ maximal supergravity,
these gaugings has been worked out in the eighties \cite{hull}, 
but some difficulties prevented up to now their construction in the 
five dimensional case. These difficulties are related to the 
feature of one--form/two--form duality, typical of five--dimensions.  
As long as all vector fields are
abelian we can consider them as one--form or two--form gauge
potentials at our own will. Yet when we introduce non--abelian gauge 
symmetry matters become more complicated,
since only $1$--forms can gauge non--abelian groups while
$2$--forms cannot. On the other hand $1$--forms that transform in
a non--trivial representation of a non abelian gauge group which
is not the adjoint representation are equally inconsistent. They
have to be replaced by $2$--forms and some other mechanism,
different from gauge symmetry has to be found to half their
degrees of freedom. This is self--duality between the $2$--form
and its field strength \cite{townsend}. 
Hence gauged supergravity can only exist
with an appropriate mixture of $1$--forms and self--dual
$2$--forms. While this mixture was mastered in the case of compact
and non--compact but semisimple gaugings, the case of
non--semisimple algebras is more involved. 
Furthermore, it appears problematic to write a Lagrangian for 
the theories with non--semisimple gaugings. Again, the problems 
come from the mixture of $1$--forms and self--dual $2$--forms.
\par
The catch of \cite{cso} is the use of the geometric approach 
(for a review of this topic see 
\cite{castdauriafre}) where the mechanism of one--form/two--form 
dualization receives a natural algebraic formulation and explanation.
The result is that in the case of the non--semisimple gaugings there
are $15+r$ gauge vectors and $12-r$ self--dual two--forms. $15$ of the
vectors gauge the non--semisimple algebra while $r$ of them have an 
abelian 
gauge symmetry with respect to which no field in the theory is charged. 
At the same time these vectors are neutral with respect to the 
transformations of the gauge algebra. 
\par
In the case of $r>0$ extra neutral vector fields, although field equations 
can be normally derived from closure of the supersymmetry algebra, a
lagrangian of conventional type might not exist, just as it
happens for type $IIB$ supergravity in $D=10$ (after all, this is
not terribly surprising since ${\cal N}=8$ supergravities in five
dimensions should eventually be interpreted in terms of brane
mechanisms and compactifications from type IIB superstring).
This would make impossible the construction of the theory
by means of lagrangian--based techniques.
However in our construction, the existence of a Lagrangian is not
fundamental, the existence of the theory following from the
consistent closure of Bianchi identities.
\par
The scalar potential of these supergravities can be systematically
derived, together with the complete field equations, from the closure of 
the supersymmetry algebra we have determined in \cite{cso}.
This is completely algorithmic and straightforward, but it involves lengthy
calculations that are postponed to a forthcoming publication \cite{next}.
\section{$D=5~~{\cal N}=8$ supergravity}
In this section we recall the main features of
$D=5~~{\cal N}=8$ supergravity theory \cite{gunwar},
\cite{PPV}.
The supersymmetry algebra for the ungauged theory is the
superPoincar\'e superalgebra, whose external automorphism
symmetry (the $\cal R$-symmetry) is $USp(8)$. The
theory is invariant under local $ISO(4,\!1)\times USp(8)$ and
global $E_{6(6)}$ transformations,
and under local supersymmetry transformations, generated by $32$ real
supersymmetry charges, organized in $8$ pseudo--Majorana spinors.
\par
The theory contains: the graviton field, namely the f\"unfbein 1--form 
$V^a$, eight gravitinos $\psi^A \equiv \psi^A_\mu \, dx^\mu$ in the
$\bf 8$ representation of $USp(8)$, $27$ vector fields 
$A^{\Lambda} \equiv A^{\Lambda} _\mu \, dx^\mu$ in the $\bf 27$ of
$E_{6(6)}$, $48$ dilatinos $\chi^{ABC}$ in the $\bf 48$ of $USp(8)$, and 
$42$ scalars $\phi$ that
parametrize the coset manifold $E_{\left(6\right)6}/USp(8)$, and
appear in the theory through the coset representative 
$I\!\!L_{\Lambda}^{~AB}(\phi)$, in the
$\bf (27,\overline{{27}})$ of $USp(8)\times E_{6(6)}$.
The local $USp(8)$ symmetry is gauged by a composite connection
built out of the scalar fields. 
\par
Let us consider the gauged theories.
In maximal supergravities, where no matter multiplets can be added,
{\it gauging} corresponds to the addition of suitable
interaction terms that turn  a subgroup ${\cal G}$ of the global
$E_{\left(6\right)6}$ duality group into a local symmetry.
This is done by means of vectors chosen among the $27$ $A^{\Lambda}$.
The $E_{\left(6\right)6}$ symmetry is broken to the normalizer of
${\cal G}$ in $E_{\left(6\right)6}$, and after this operation
the new theory has a local symmetry $USp(8)\!\times\!{\cal G}$
and a global symmetry $N({\cal G},E_{\left(6\right)6})$.
From group theoretical considerations one can derive necessary conditions 
which strictly constrain the choice of ${\cal G}$. Such conditions 
are satisfied only by the semisimple groups $SO(p,q)$ with
$p+q=6$ and their non--semisimple contractions $CSO(p,q,r)$,
which will be discussed in section $5$ (see
\cite{hull,noi4D} for definitions). The possible gaugings
are then restricted to these groups. The normalizer in
$E_{\left(6\right)6}$ of all these groups is $SL(2,\!I\!\!R)$. 
Therefore this latter is the residual global symmetry for all 
possible gaugings. The 27 vectors
$A^{\Lambda}$ are then decomposed into the vectors $A_{IJ}$ in the
$\bf (\bar{15},1)$, that gauge ${\cal G}$, and the vectors  in
the $\bf (6,2)$, which do not gauge anything. In the $SO(p,q)$ gaugings, 
which have been built in \cite{gunwar,PPV}, the latter $12$ vectors have 
to be dualized into two--forms $B^{I\alpha}$, as we will explain in 
the following. 
For all the admissible cases, the fifteen generators $G^{IJ}$ of ${\cal G}$
have, in the fundamental $\bf 6$--dimensional representation, the form
\begin{equation}
(G^{IJ})^K_{~L}=\delta^{[I}_L\eta^{J]K}
\label{genfund}
\end{equation}
where $\eta^{JK}$ is a diagonal matrix with $p$ eigenvalues equal to $1$, 
$q$ eigenvalues equal to $(-1)$  and, only in the case of contracted 
groups,  $r$  null eigenvalues. This signature completely characterizes
the gauge groups and correspondingly the gauged theory.
The covariant derivative with respect to ${\cal G}$
of a field $V^I$ in the $\bf 6$ of $SL(6, I\!\!R)$ is defined as
\begin{equation}
DV^I\equiv {\cal D} V^I+g(G^{KL})^I_{~J}A_{KL}\wedge V^J\,.
\label{covGder}
\end{equation}
where ${\cal D}$ is the Lorentz--covariant exterior derivative.
\par
The field content of the (semisimple) gauged supergravity theories is the 
following:
\begin{equation}
\begin{array}{|c|c|c|c|c|}
\hline
\#& \hbox{Field} & \left(SU(2)\times SU(2)\right) \hbox{--spin~rep.} & 
USp(8) \hbox{~rep.} & {\cal G}
\hbox{~rep.} \\
\hline
1& V^a & (1,1) & {\bf 1} & {\bf 1} \\
\hline
8& \psi^A & (1,1/2)\oplus (1/2,1) & {\bf 8} & {\bf 1} \\
\hline
15& A_{IJ} & (1/2,1/2) & {\bf 1} & {\bf 15} \\
\hline
12& B^{I\alpha} & (1,0)\oplus (0,1) & {\bf 1} & {\bf 6\oplus 
\overline{{6}}} \\
\hline
48& \chi^{ABC} & (1/2,0)\oplus (0,1/2) & {\bf 48} & {\bf 1} \\
\hline
42& I\!\!L_{\Lambda}^{~AB}\left(\phi\right) & (0,0) & {\bf 27} & 
{\bf\overline{{27}}} \\
\hline
\end{array}
\label{gaugedfieldcontent}
\end{equation}
\section{Gauged supergravities from 
F.D.A.'s  and Rheonomy} 
Gauged maximal supergravities in $D=5$ were originally constructed within 
the framework of  No\"ether coupling and component formalism
\cite{gunwar},\cite{PPV}. As we pointed out, 
the gaugings corresponding to the contracted groups
$CSO(p,q,r)$ were left open in that approach. 
\par
In \cite{cso}, these theories have been constructed by using the
approach based on free differential algebras (F.D.A.'s) and rheonomy. 
To obtain this result, we started by reformulating the $D=5$ ${\cal N}=8$ 
supergravities \cite{gunwar} \cite{PPV} with semisimple gauge groups 
in the rheonomic framework \cite{castdauriafre}. 
Then, the extension to non--semisimple gaugings has been worked out. 
\par
Here we do not describe the rheonomic formulation of 
supergravity. For a comprehensive review, see \cite{castdauriafre}, 
while for a short summary, in the context of $D=5$ ${\cal N}=8$ 
supergravity, see \cite{cso}.
We only recall that this approach is based on the closure of 
the Bianchi Identities of the superspace curvatures. 
In this context, the Bianchi Identities 
are not identically satisfied. Actually, they are the 
equations of the theory, determining its dynamics. Not only they give the
parametrizations of the curvatures (from which one can read off the 
supersymmetry transformations), but they also fix the geometry of the 
scalar manifold and give the classical field equations satisfied by the 
spacetime fields.
From this viewpoint, the explicit construction of the Lagrangian $\cal L$ 
is not really needed. Simply, when $\cal L$ exists, the determination of 
the field equations is more easily obtained by $\delta{\cal L}$ variations 
than through the analysis of the Bianchi Identities.
When the Lagrangian exists, it can be obtained by means of a 
straightforward procedure starting from the curvature parametrizations 
\cite{castdauriafre}.
\par
This construction has been performed in \cite{cso}, by solving the 
Bianchi Identities of $D=5$ ${\cal N}=8$ supergravity, both for the 
semisimple and non--semisimple cases.
The parametrizations of the curvatures, and then the supersymmetry 
transformations, have been determined (modulo bilinear in the dilatinos). 
In the semisimple case, they are 
\begin{eqnarray}
\delta V^a_{\mu} &=& - {\rm i} \bar{\varepsilon}^A \gamma^a\psi_{\mu A}  \\
\delta \psi_{A \mu}  &=&{\cal D}_\mu \varepsilon_A - g
{2\over 45}T_{AB}\gamma_\mu \varepsilon^{B} +  {2\over 3}
{\cal H}_{AB|\nu\mu} \gamma^{\nu} \varepsilon^B -{1\over
12}{\cal H}_{AB}^{\nu\rho} \gamma^{\lambda\sigma}
\varepsilon^B \epsilon _{\mu\nu\rho\lambda\sigma} \nonumber\\\nonumber\\
&& +{3{\rm i}\over 2\sqrt{2}}
\chi_{ABC}\bar{\varepsilon}^{B}\psi_{\mu}^C - {{\rm i}\over
2\sqrt{2}} \gamma_\nu\chi_{ABC} \bar\varepsilon^B\gamma^\nu
\psi_\mu^C + \mathcal{O} (\chi^2)  \\\nonumber\\
\delta \chi_{ABC} &=& {1\over\sqrt{2}} gA^D_{ABC}\varepsilon_D
+\sqrt{2} \hat{P}_{ABCD|i}\partial_\nu\phi^i
\gamma^\nu\varepsilon^D - {3\over 2\sqrt{2}}
{\cal H}_{[AB|\mu\nu}\gamma^{\mu\nu}\varepsilon_{C]} \nonumber\\\nonumber\\
&& - \frac{1}{2\sqrt{2}}
\Omega_{[AB}{\cal H}_{C]D|\mu\nu}\gamma^{\mu\nu}\varepsilon^D +
{\cal O} (\chi^2) \\\nonumber\\
\delta A_{IJ|\mu} &=& I\!\!L^{-1}_{ABIJ}\left[ {{\rm
i}\over\sqrt{2}} \bar{\chi}^{ABC}\gamma_\mu\varepsilon_C
+2{\rm i}\bar\varepsilon^A\psi^B_\mu \right]\\\nonumber\\
\delta B^{I\alpha}_{\mu\nu} &=& I\!\!L^{I\alpha}_{~~AB}\left[
- 2{\rm i} g \bar{\varepsilon}^A\gamma_{[\mu}\psi^B_{\nu]}
- {{\rm i}\over 2\sqrt{2}}g\bar{\chi}^{ABC}\gamma_{\mu\nu}
\varepsilon_C\right]\nonumber\\\nonumber\\
&&+2{\cal D}_{[\mu}\left[ I\!\!L^{-1~I\alpha}_{AB}\left(
2{\rm i}\bar\varepsilon^A \psi^B_{\nu]} +{{\rm i}\over\sqrt{2}}
\bar{\chi}^{ABC}\gamma_{\nu]}\varepsilon_C\right)\right]
\label{beq}
 \\\nonumber\\
\hat{P}^{ABCD}_{,i}\delta \phi^i &=& 2{\rm i}\sqrt{2}\,
\bar{\chi}^{[ABC}\varepsilon^{D]}+\frac{3{\rm
i}}{\sqrt{2}}\,\Omega^{[CD}\chi^{AB]E}\varepsilon_E
\end{eqnarray}
as in \cite{gunwar}. In the semisimple case, 
also the Lagrangian and the equations of motion have been determined, 
and found to coincide with the results of \cite{gunwar}.
We postpone to section \ref{seccso} the analysis of the non-semisimple 
case.
\section{The problem of the two--forms}
It is a known fact \cite{PPV}, \cite{gunwar} that in order to
consistently gauge the ${\cal N}=8$ theory, one has to dualize
the vectors transforming in the $(\bf{6},\bf{2})$ of
$SO(p,q)\times SL(2,I\!\!R)$ to massive two-forms
obeying the self-duality constraint:
\begin{equation}
B^{I\alpha |\mu\nu} = m
\epsilon^{\mu\nu\rho\sigma\lambda}{\cal D}_\rho
B^{I\alpha}_{\rho\sigma\lambda} \label{self}
\end{equation}
with $m\sim g$. In the geometric formulation of the theory, this need for
dualization emerges in a completely natural way. Indeed, let us
start by considering the 12 vectors $A^{I\alpha}$. There is no way
known to reconcile their abelian gauge invariance with their
non-trivial transformation under the gauge group $\mathcal{G}$.
Indeed, given the superspace curvatures
\begin{equation}
DA^{I\alpha} \equiv dA^{I\alpha}+g(G^{KL})^I_{\,J}A_{KL}\wedge A^{J\alpha}
\end{equation}
it follows that the corresponding Bianchi Identities contain a term
\begin{equation}
DDA^{I\alpha} = g(G^{KL})^I_{\,J}{\cal F}_{KL}\wedge
A^{J\alpha}
\end{equation}
where the vectors $A^{J\alpha}$ appear naked. This makes impossible 
to write a parametrization of the curvatures covariant under the 
gauge group. Hence we have a clash between supersymmetry
and the $12$ abelian gauge invariances needed to keep the vectors
$A^{J\alpha}$ massless. On the other hand, making them massive
would destroy the equality of the Bose and Fermi degrees of
freedom. So in the gauged case, where the $12$ vectors
$A^{J\alpha}$ acquire a non--trivial transformation under the
non--abelian gauge symmetry, there is no way of fitting these
fields into a  consistent supersymmetric theory. The way out, as
it was discussed in \cite{gunwar}, is to interpret them as the
duals of massive two-forms $B^{I\alpha}$, 
obeying a self--duality constraint which halves
their degrees of freedom. This construction emerges naturally in
the rheonomic framework \cite{cso}. In
this context, one has to introduce superspace curvatures for the
two--forms generalizing the
Maurer--Cartan equations to a F.D.A.
\cite{castdauriafre,Fre:1984pc}. At first sight it seems
that we cannot escape from the problem described above, that
affects the vectors $A^{I\alpha}$:  indeed Bianchi identities do
contain the naked fields $B^{I\alpha}$. Yet we can successfully
handle this fact by considering the $B^{I\alpha}$ not as gauge
potentials (that is, 2-forms defined modulo $1$--form gauge
transformations), but as physical fields, with their own explicit
parametrization \footnote{The same
happens to matter two--form fields coupled with ${\cal N}=2$
supergravity \cite{Ceresole:2000jd}.}. In this way, the two--forms
loose their gauge freedom and become massive, as it can be found by 
solving the Bianchi identities. In fact, the Bianchi identities give 
directly the self--duality constraint (\ref{self}) obeyed by the 
two--forms:
\begin{equation}
\label{eqmotB} 
D_{[a}B^{I\alpha}_{bc]}= -{1\over 12} g
I\!\!L^{I\alpha}_{~~AB}{\cal H}^{AB~|de}\varepsilon_{abcde}
+{\rm fermion\,\,terms}\,.
\end{equation}
\section{Gauging the non--semisimple $CSO(p,q,r)$ groups}
\label{seccso}
Let us consider the gauging of the $\!CSO(p,q,r)$ $p\!+\!q\!+\!r\!=\!6$
groups in $D\!=\!5$ ${\cal N}\!=\!8$ supergravity. 
We begin with a short description of the $CSO(p,q,r)$ algebras
(see also \cite{hull,noi4D}).
The generators of $SO(p,q)$ (with $p+q=n$) satisfy
\begin{equation}
[G^{IJ},G^{KL}]=f^{IJ,KL}_{MN}G^{MN}
\end{equation}
where
\begin{equation}
\label{structconst}
f^{IJ,KL}_{MN}=-2\delta^{[I}_{[M}\eta^{J][K}\delta^{L]}_{N]}
\end{equation}
and
\begin{equation}
\eta^{IJ}\equiv\,{\rm diag}
(\overbrace{1,\dots,1}^p,\overbrace{-1,\dots,-1}^q)\,.
\label{metricpq}
\end{equation}
Their generalization, studied by Hull in the context of
supergravity \cite{hull} are the algebras
$CSO(p,q,r)$ with $p+q+r=n$, defined by the structure
constants (\ref{structconst}) with
\begin{equation}
\eta^{IJ}\equiv\,{\rm
diag}(\overbrace{1,\dots,1}^p,\overbrace{-1,\dots,-1}^q,
\overbrace{0,\dots,0}^r)\,. \label{metriccpqr}
\end{equation}
Decomposing the indices as 
$I\!=\!(\bar{I},\hat{I})~~\bar{I}\!=\!1,\!\dots\!,p\!+\!q,
~\hat{I}=\!p\!+\!q\!+\!1,\!\dots\!,n\,,$
we have that $G^{\bar{I}\bar{J}}$ are the generators of
$SO(p,q)\!\subset\!CSO(p,q,r)$ , while the
$r(r\!-\!1)/2\,\,$ $G^{\hat{I}\hat{J}}$ are central charges
\begin{equation}
[G^{\bar{I}\hat{J}},G^{\bar{K}\hat{L}}]={1\over
2}\eta^{\bar{I}\bar{K}} G^{\hat{J}\hat{L}}\,. \label{nonnull}
\end{equation}
They form an abelian subalgebra, and
\begin{equation}
SO(p,q)\times U(1)^{r\left(r-1\right)\over
2}\subset CSO(p,q,r)\,.
\end{equation}
As in the case of $SO(p,q)$ (see eq. (\ref{genfund})), the 
\begin{equation}
(G^{IJ})^K_{~~L}=\delta^{[K}_J\eta^{L]I}~~~~I,J,K,L=1,\dots,n
\label{genvect}
\end{equation}
are generators of a representation of $CSO(p,q,r)$. This representation 
is not faithful, because the generators 
of the central charges are identically null
\begin{equation}
(G^{\hat{I}\hat{J}})^K_{~L}=0\,. \label{zerogen}
\end{equation}
\par
The gauged versions of ${\cal N}=8$, $D=5$
supergravity constructed in \cite{gunwar}, \cite{PPV} and based on
a semisimple choice of the gauge group
${\cal G}=SO(p,q)~(p\!+\!q\!=\!6)$ can be 
generalized (see \cite{cso}) to the non--semisimple gauge groups
${\cal G}=CSO(p,q,r)~(p\!+\!q\!+\!r\!=\!6)$.
\par
The new gaugings can be obtained by taking for the matrix
$\eta^{IJ}$ the definition (\ref{metriccpqr}), with some null
entries on the diagonal. Let us discuss the consequences of  this
in the theory, in order to see if any pathology occurs. One has
\begin{equation}
(G^{KL})^{\hat{I}}_{~J}=\delta^{[K}_J\eta^{L]\hat{I}}=0\,,
\end{equation}
so the covariant derivative of a {\sl contravariant field}
(\ref{covGder}), along the contracted directions, reduces to the
ordinary  Lorentz--covariant derivative:
\begin{equation}
DV^{\hat{I}}= {\cal D}
V^{\hat{I}}+g(G^{KL})^{\hat{I}}_{~J}A_{KL}\wedge V^J ={\cal D}
V^{\hat{I}}\,.
\label{discharged}
\end{equation}
This, however, does not happen for the covariant derivative of a
{\sl covariant field}:
\begin{equation}
DV_{\hat{I}} \equiv \nabla
V_{\hat{I}}-g(G^{KL})^J_{~\hat{I}}A_{KL}\wedge V^J= \nabla
V_{\hat{I}}-g\eta^{\bar{L}\bar{J}}A_{\hat{I}\bar{L}}\wedge
V_{\bar{J}}\,.
\end{equation}
Let us consider now the most subtle part of the theory:
the two--forms. First of all we notice that, because of 
(\ref{discharged}), the two--forms along the contracted directions 
$B^{\hat{I}\alpha}$ are discharged under the gauge group. 
Furthermore, one finds that the Bianchi identities of the two--forms
corresponding to the contracted directions (the
$B^{\hat{I}\alpha}$) are cohomologically trivial, so that these
fields are actually field strengths of one--form fields
\begin{equation}
B^{\hat{I}\alpha}\equiv dA^{\hat{I}\alpha}+{\rm\,\,fermions}
\label{bianchiAhatI}
\end{equation}
having a $U(1)$ gauge
invariance, as argued in \cite{gunwar}. Let us stress that the
calculation of \cite{cso} shows
that there are no consistency conflicts between the two types of
gauge invariances, and therefore no need arises to introduce
massive vectors as proposed in \cite{gunwar}. Indeed,  in the F.D.A.
rheonomic approach we  see in a transparent way where the
consistency conflicts arise and how they are solved. Summarizing
it goes as follows. When a vector field is charged with respect to
the gauge group, but does not gauge any generator of the gauge
algebra it appears naked in its own Bianchi identities. This
requires dualization to a two--form, so the correct number of degrees 
of freedom is got through self--duality instead of gauge invariance.
On the other hand  when a contraction is
performed on some direction $\hat{I}$, in the Bianchi identities of
the fields $A^{\hat{I}\alpha}$ (\ref{bianchiAhatI}) the naked
gauge fields disappear. Therefore, the two gauge invariances are
not inconsistent, and the corresponding vectors can stay massless.
Note that in this case the Bianchi identities look very different
from those along the non-contracted directions. Now the
self--duality constraint disappears and the halving of degrees of
freedom is due to the recovered $U(1)$ gauge symmetry.
\par
In this way new gauged $D=5~{\cal N}=8$
supergravities arise, with $(12-r)$ two--forms, $(15+r)$ one--forms, and
gauge group $CSO(p,q,r)$. It is worth noting
that the $r$ vectors $A^{\hat{I}\alpha}$ are coupled with the
other fields, even if they don't gauge anything, and so are the
abelian  vectors $A_{\hat{I}\hat{J}}$. Indeed,
\begin{equation}
{\cal H}^{AB}_{ab}=I\!\!L^{IJAB}F_{ab\vert IJ}+
I\!\!L_{\bar{I}\alpha}^{~~AB}B^{\bar{I}\alpha}_{ab}+
I\!\!L_{\hat{I}\alpha}^{~~AB}B^{\hat{I}\alpha}_{ab}
\end{equation}
and ${\cal H}^{AB}_{ab}$ does appear in the equations of motion
of the two--forms (\ref{eqmotB}) along the non--contracted
directions, which we have derived from the Bianchi identities 
and don't change in the contracted gaugings.
\par
The supersymmetry transformation rules for the new theories are obtained 
by substituting, for the contracted directions $\hat I$, the supersymmetry 
transformation rule (\ref{beq}) 
for $B^{\hat{I}\alpha}$ with the supersymmetry transformation for 
$A^{\hat{I}\alpha}$ 
\begin{equation}
\delta A^{\hat{I}\alpha}_{\mu} = I\!\!L^{-1~\hat
I\alpha}_{AB}\left[ {{\rm i}\over\sqrt{2}}
\bar{\chi}^{ABC}\gamma_\mu\varepsilon_C +2{\rm
i}\bar\varepsilon^A\psi^B_\mu \right]
\end{equation}
all other transformation laws remaining unchanged.
The new theories are completely sensible and well
defined, however apparently there is not a lagrangian formulation
of them. In fact, it seems impossible to write the terms giving the 
equations of motion of the $r$ one--form fields $A^{\hat{I}\alpha}$ 
in a covariant and gauge--invariant way. A possible argument to
motivate this situation is the following. 
The existence of $r$ extra neutral vectors besides
the $15$ charged ones implies a sort of Hodge dualization for the
corresponding two--forms. Specifically, what happens here is that
the field strength $H^{[3]}$ for $r$ of the $B^{[2]}$ fields is
identically zero, so that we have to interpret the $B^{[2]}$
themselves as field strengths of new gauge vectors $A^{[1]}$. In
other words, we have traded $r$ ''electric'' two--form fields
$B^{[2]}$ for just as many ''magnetic''  one--forms $A^{[1]}$. In
view of this, it is not too surprising if the $15+r$ vectors are
not mutually local, which would be necessary to admit a common
lagrangian description.
\vskip0.5cm
\noindent
{\large \bf Acknowledgements}
\smallskip
\noindent
L.G. is supported in part by the ``Actions de Recherche Concert{\'e}es"
of the ``Direction de la Recherche Scientifique - Communaut{\'e}
Fran{\c c}aise de Belgique", by IISN - Belgium (convention
4.4505.86). This work has been supported by the European
Commission RTN programme HPRN-CT-2000-00131 in which L.A. and P.F.
are associated to Torino and L.G. is associated to K.\ U.\ Leuven.
These proceedings have been written while participating in the programme 
{\it Supergravity, Superstrings and $M$--Theory} (Sept. 18, 2000 - 
Feb. 9, 2000) of the Centre \'Emile Borel at the Institut Henri Poincar\'e 
(UMS 839 - CNRS/UPMC), during which L. G. was funded by a C.N.R.S. 
support obtained by the Centre \'Emile Borel.

\end{document}